\documentclass[aps,pre,amsmath,10pt,twocolumn,superscriptaddress]{revtex4-1}
\usepackage{graphicx}
\usepackage{bm}
\usepackage{color}

\begin{document}
\title{Prediction of perovskite-related structures in ACuO$_{3-x}$ (A $=$ Ca, Sr, Ba, Sc, Y, La) using density functional theory and Bayesian optimization}
\author{Atsuto \surname{Seko}}
\email{seko@cms.mtl.kyoto-u.ac.jp}
\affiliation{Department of Materials Science and Engineering, Kyoto University, Kyoto 606-8501, Japan}
\affiliation{Center for Elements Strategy Initiative for Structure Materials (ESISM), Kyoto University, Kyoto 606-8501, Japan}
\affiliation{Center for Materials Research by Information Integration, National Institute for Materials Science, Tsukuba 305-0047, Japan}
\author{Shintaro \surname{Ishiwata}}
\affiliation{Division of Materials Physics, Graduate School of Engineering Science, Osaka University, Toyonaka, Osaka 560-8531, Japan}

\date{\today}

\begin{abstract}
Oxygen vacancy ordering in perovskite-type transition-metal oxides plays an important role in the emergence of exotic electronic properties, as typified by superconducting cuprates.
In this study, we predict the stability of oxygen-deficient perovskite structures in ACuO$_{3-x}$ (A $=$ Ca, Sr, Ba, Sc, Y, La) by density functional theory calculation.
We introduce a combination of the cluster expansion method, Gaussian process, and Bayesian optimization to find stable oxygen-deficient structures among a considerable number of candidates.
Our calculations not only reproduce the reported structures but suggest the presence of several unknown oxygen-deficient perovskite structures, some of which are stabilized at high pressures. 
This work demonstrates the great applicability of the present computational procedure for the elucidation of the structural stability of strongly correlated oxides with a large tolerance to oxygen deficiency.
\end{abstract}

\maketitle

\section{Introduction}

Perovskite-type transition-metal oxides, namely, ABO$_3$, have been extensively studied as strongly correlated systems showing a large variety of electronic properties, typified by high-$T_c$ cuprates and ferromagnetic manganites. 
One of the major methods of controlling the physical properties of ABO$_3$ is the chemical substitution for A-site cations, which dominates the bandwidth and valence of B-site cations \cite{doi:10.1021/cm00026a003,B512271F,tilley2016perovskites}. 
Their potential to provide novel electronic properties can be further enhanced by introducing oxygen deficiency, which plays an important role not only in the valence of B-site cations but also in the topology of the B--O lattice.
For instance, the typical superconducting cuprates YBa$_2$Cu$_3$O$_{7-d}$ \cite{PhysRevLett.58.1676} and Sr$_{1-x}$La$_x$CuO$_2$ \cite{doi:10.7566/JPSJ.82.063705} are well-ordered oxygen-deficient perovskites with CuO$_2$ planes. 
In addition, SrCuO$_{2.5}$ \cite{CHEN1996498} and LaCuO$_{3-x}$ with $x=0.4$ and 0.5 \cite{bringley1990synthesis} have been reported as potential superconducting cuprates with a quasi-one-dimensional Cu--O motif.


Considering the recent development of computational structure prediction methods and computational resources, it is worth performing a comprehensive search for oxygen-deficient perovskite-type structures and related ones to discover novel cuprates that have been overlooked so far. 
Recently, computational studies on perovskite-type transition-metal oxides, ABO$_{3-x}$, have been reported, but their structural variations are quite limited \cite{emery2017high}.
In this study, we systematically investigate the stability of oxygen-deficient perovskite structures in ACuO$_{3-x}$ ($0 \leq x \leq 1$) with divalent and trivalent cations (A $=$ Ca, Sr, Ba, Sc, Y, La) by density functional theory (DFT) calculation.  
This study is based on a large set of oxygen-vacancy configurations derived from the cubic perovskite lattice to search for a stable structure with a large unit cell volume.
Enumeration of the nonequivalent configurations is shown in Sec. \ref{perovskite:section-enum}.

A common way to find stable structures in such a configurational system is the cluster expansion (CE) method \cite{CE1,CE2,CE3}.
The CE method has been successfully used to describe alloy thermodynamics and to efficiently find stable alloy configurations among a considerable number of candidates.
Although the CE method has also been employed to predict atom-deficient nonstoichiometric structures \cite{PhysRevB.58.2975,PhysRevLett.87.275508,PhysRevB.66.064112,SnOx:Seko,PhysRevB.77.144104}, it fails to accurately represent the configurational energy in oxygen-deficient ACuO$_{3-x}$ with large geometry relaxation from the ideal cubic perovskite lattice, as shown in Sec. \ref{perovskite:section-ce}.
Therefore, we introduce Bayesian optimization \cite{jones01}, which has recently been applied to the exploration of materials and structure prediction \cite{seko2014machine,PhysRevLett.115.205901,PhysRevB.93.054112,xue2016accelerated,Kiyoharae1600746,PhysRevMaterials.2.013803,PhysRevB.97.125124,todorovic2019bayesian}.
As shown in Sec. \ref{perovskite:section-bo}, Bayesian optimization with the Gaussian process (GP) \cite{Rasmussen_2006} and structural representations of the CE method is very useful for predicting oxygen-deficient structures.

\section{Nonequivalent oxygen-deficient perovskite structures}
\label{perovskite:section-enum}

\begin{figure}[tbp]
\includegraphics[clip,width=\linewidth]{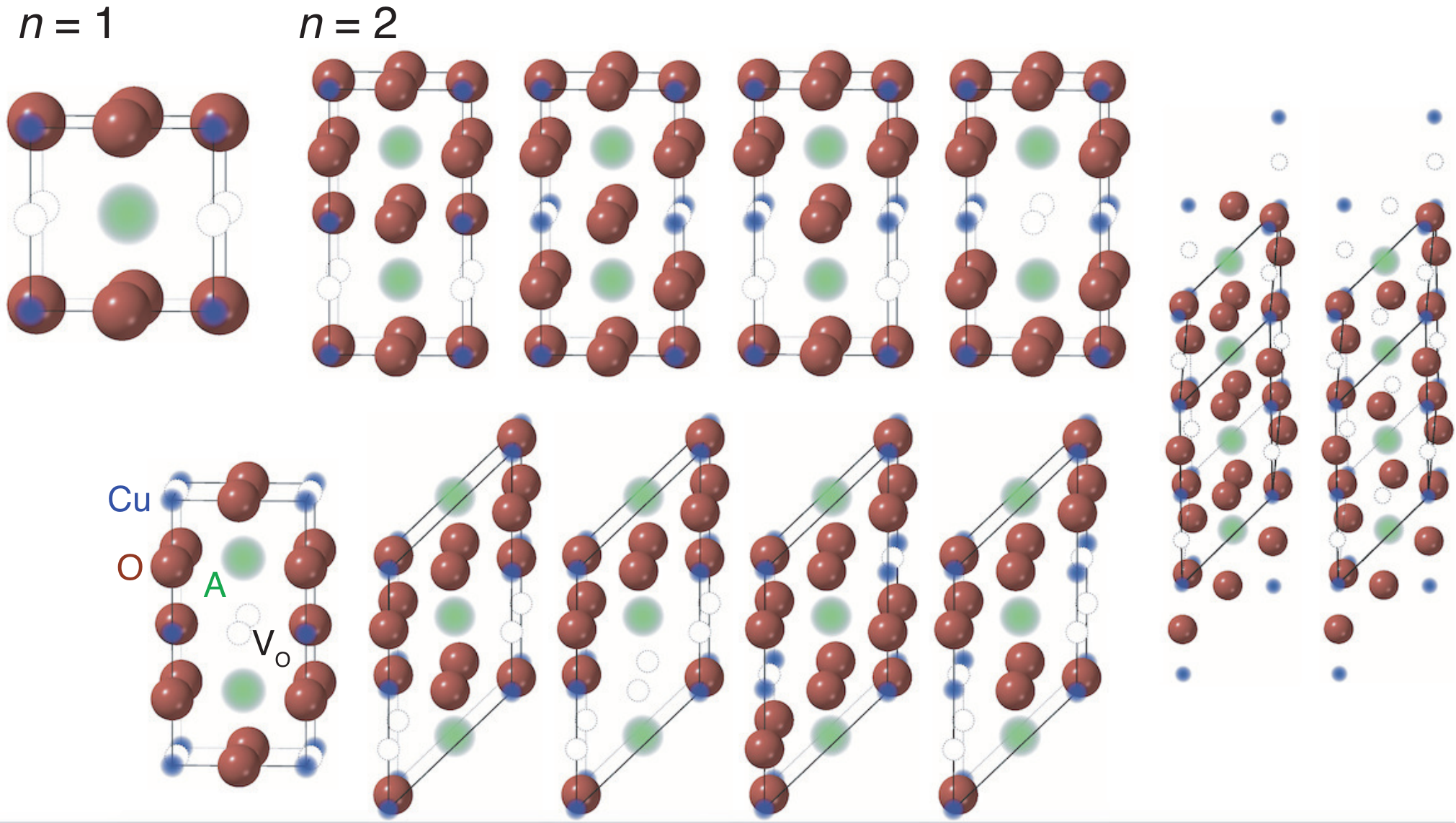}
\caption{
A complete set of nonequivalent oxygen-deficient structures of cubic perovskite for supercell sizes of $n=1$ and $n=2$ in ACuO$_{3-x}$ ($0 \leq x \leq 1$).
}
\label{perovskite:structure-example}
\end{figure}

A pool of oxygen-deficient structures with the cubic perovskite lattice is firstly generated using a derivative structure algorithm proposed by Hart and Forcade \cite{Hart_derivativestructure,Hart_derivativestructure2} implemented in the \textsc{clupan} code \cite{sampling:seko,clupan}. 
The derivative structure algorithm consists of two steps of enumeration.
In the first step, a complete set of nonequivalent supercell shapes of the cubic perovskite is enumerated for a given supercell size $n$. 
The supercell shape is identified by the three-dimensional square Hermite normal form (HNF), $\bm{H}$, with the determinant $n = |\bm{H}|$, modifying the axis matrix of the conventional unit cell $\bm{A}$ as
\begin{equation}
\bm{A'} = \bm{AH}.
\end{equation}
Therefore, the enumeration of nonequivalent supercell shapes corresponds to that of nonequivalent HNFs.
For instance, a complete set of nonequivalent HNFs for up to $n=8$ is listed in Appendix (Table \ref{perovskite:list-hnf}).
In the second step, nonequivalent oxygen-deficient structures are enumerated for each HNF using the finitely presented group of the supercell lattice constructed by the HNF.
Figure \ref{perovskite:structure-example} illustrates a complete set of nonequivalent oxygen-deficient structures with supercell sizes of $n=1$ and $n=2$.
Moreover, Table \ref{perovskite-2019:n-derivative} lists the number of nonequivalent oxygen-deficient structures in ACuO$_{3-x}$ ($0 \leq x \leq 1$).
The number of nonequivalent structures increases exponentially as the supercell size increases.

\begin{table}[tbp]
\begin{ruledtabular}
\caption{
Number of nonequivalent oxygen-deficient structures of cubic perovskite in ACuO$_{3-x}$ ($0 \leq x \leq 1$).
Symbols $N_{\rm sites}$, $N_{\rm config}$, and $N_{\rm config, acc}$ denote the number of oxygen lattice sites, the number of nonequivalent configurations for given $n$, and the cumulative number of nonequivalent configurations with up to given $n$, respectively.
}
\label{perovskite-2019:n-derivative}
\begin{tabular}{cccc}
$n$ & $N_{\rm sites}$ & $N_{\rm config}$ & $N_{\rm config, acc}$ \\
\hline
1 & 3  & 1      & 1 \\
2 & 6  & 11     & 12 \\
3 & 9  & 42     & 54 \\
4 & 12 & 444    & 498 \\
5 & 15 & 1245   & 1743 \\
6 & 18 & 16343  & 18086 \\
7 & 21 & 49403  & 67489 \\
8 & 24 & 706622 & 774111 \\
\end{tabular}
\end{ruledtabular}
\end{table} 

\section{Cluster expansion method}
\label{perovskite:section-ce}
\subsection{Formalism}
In general, it is impossible to estimate the configurational energy for a complete set of nonequivalent structures by DFT calculation owing to the considerable number of nonequivalent structures. 
Therefore, it has been a practical way to employ a machine learning model of the configurational energy estimated from a set of DFT calculations for sampled structures.
The CE method \cite{CE1,CE2,CE3} has been a widely used approach, and gives a reliable model of configurational energy; hence, it has enabled us to accurately predict the ground-state structures and finite-temperature thermodynamics in many multicomponent systems.

In a binary system, the configurational energy $E$ is described in a linear form using the pseudospin configurational variable $\sigma_i$ for the respective lattice site $i$ as
\begin{eqnarray}
E &=& V_0 + \sum\limits_{i} {V_i \sigma_i} + \sum\limits_{i,j} {V_{ij} \sigma_i \sigma_j} + \sum\limits_{i,j,k} {V_{ijk} \sigma_i \sigma_j \sigma_k} + \cdots \nonumber \\
    &=& \sum\limits_{\alpha} V_\alpha \cdot \varphi_\alpha ,
\label{ce-hamiltonian}
\end{eqnarray}
where $ \varphi_\alpha $ denotes the correlation function of cluster $\alpha$.
Coefficients $V_\alpha$ are called the effective cluster interactions, which are estimated from a set of DFT calculations for sampled configurations using general linear regression methods such as conventional linear regression and the least absolute shrinkage and selection operator (LASSO) \cite{nelson2013compressive,hastieelements}.

\subsection{Application to oxygen-deficient perovskite}

\begin{figure}[tbp]
\includegraphics[clip,width=\linewidth]{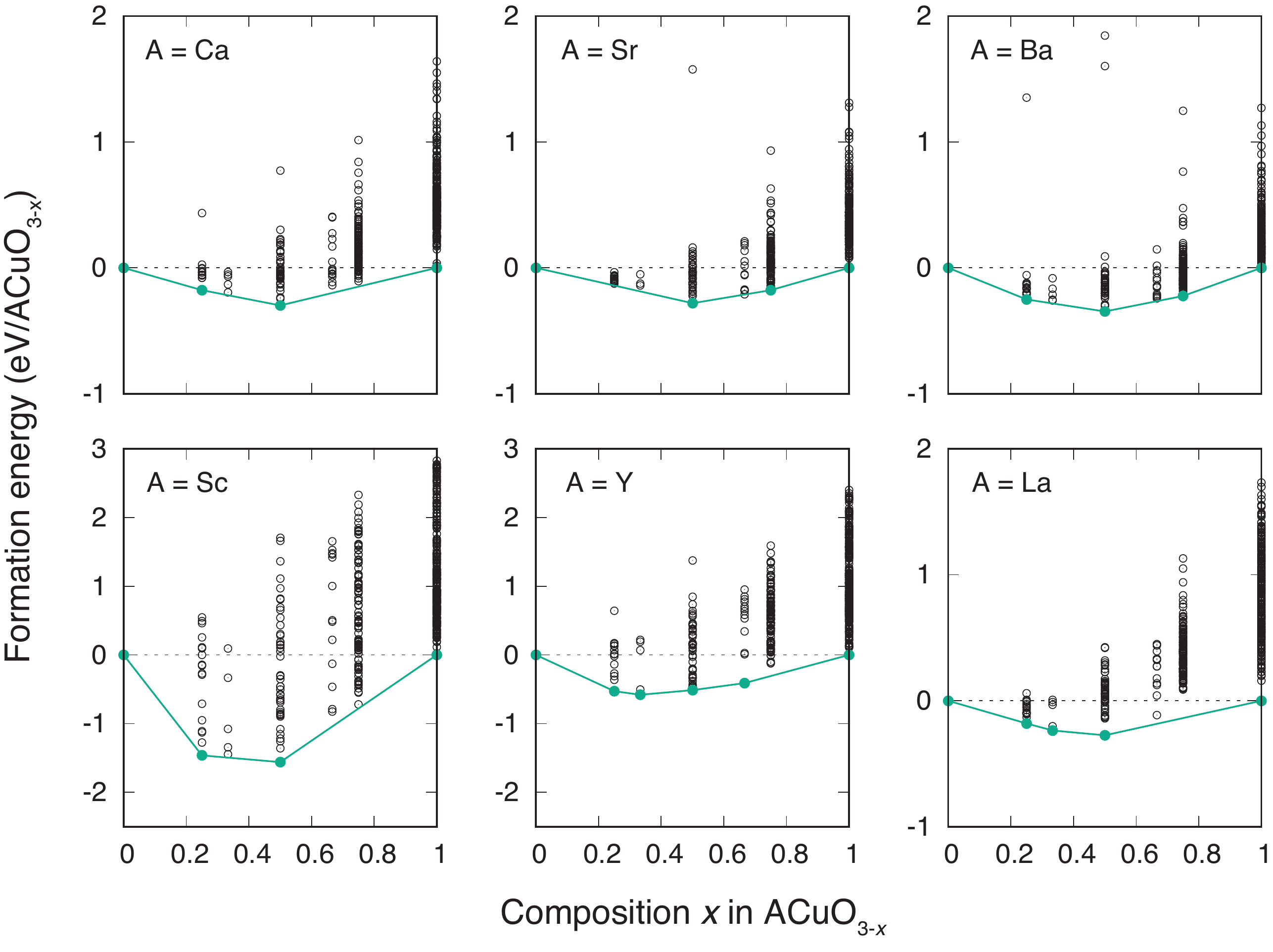}
\caption{
Formation energies of nonequivalent oxygen-deficient perovskite structures with up to $n=4$. 
The green line shows the convex hull of formation energy.
}
\label{perovskite:energy-4}
\end{figure}

We prepared a training dataset that contains the formation energies of all nonequivalent structures with up to $n=4$ estimated by DFT calculation performed using the plane-wave-basis projector augmented wave (PAW) method \cite{PAW1} within the Perdew--Burke--Ernzerhof generalized gradient approximation (GGA) exchange-correlation functional \cite{GGA:PBE96}, as implemented in the \textsc{vasp} code \cite{VASP1,VASP2,PAW2}.
The cutoff energy was set to 400 eV.
The total energies converged to less than 10$^{-3}$ meV/supercell.
The atomic positions and lattice constants of the nonequivalent structure were optimized until the residual forces were less than 10$^{-2}$ eV/\AA\ to consider the relaxation effect of the ideal cubic perovskite lattice.
We assumed the ferromagnetic configuration for copper atoms in the DFT calculation for simplicity, although ACuO$_{3-x}$ is expected to show an antiferromagnetic configuration as shown experimentally in Ca$_{1-x}$Sr$_x$CuO$_2$ \cite{PhysRevB.39.9122}.
We adopted the $+U$ approach to improve the description of the ground state of correlated systems \cite{PhysRevB.57.1505}.
The value of $U$ for copper was set to 4 eV taken from the literature \cite{PhysRevB.73.195107}.

Figure \ref{perovskite:energy-4} shows the formation energies of nonequivalent oxygen-deficient structures with up to $n=4$ in ACuO$_{3-x}$ systems. 
They are measured from the energy of the cubic perovskite ACuO$_3$ and the lowest energy among the structures in ACuO$_2$.
Since the DFT calculation fails to finish successfully in some structures, we exclude the structures from the dataset.
The sizes of the resultant training datasets are 485, 473, 480, 457, 468, and 480 for BaCuO$_{3-x}$, CaCuO$_{3-x}$, SrCuO$_{3-x}$, ScCuO$_{3-x}$, YCuO$_{3-x}$, and LaCuO$_{3-x}$, respectively.

\begin{table}[tbp]
\begin{ruledtabular}
\caption{
Fitting errors for four CE models (Unit: eV/ACuO$_{3-x}$). 
The fitting error normalized by the absolute value of the lowest formation energy is also shown in parentheses.
$N_{c}$ denotes the number of regression coefficients included in a CE model.
Models 1, 2, 3, and 4 are composed of 11 pairs, 19 pairs, 11 pairs and 23 multiplets of up to fourbodies, and 11 pairs and 39 multiplets of up to sixbodies, respectively. 
All of them also contain the empty and point clusters corresponding to the bias term and the composition $x$, respectively.
}
\label{perovskite:error-ce}
\begin{tabular}{ccccc}
& Model 1 & Model 2 & Model 3 & Model 4 \\
\hline
CaCuO$_{3-x}$ & 0.171  & 0.167 & 0.157 & 0.152 \\
 & (0.572) & (0.559) & (0.525) & (0.508) \\
SrCuO$_{3-x}$ & 0.135  & 0.133 & 0.127 & 0.125 \\
 & (0.480) & (0.473) & (0.451) & (0.445) \\
BaCuO$_{3-x}$ & 0.170  & 0.169 & 0.159 & 0.155 \\
 & (0.491) & (0.488) & (0.459) & (0.448) \\
ScCuO$_{3-x}$ & 0.669  & 0.657 & 0.614 & 0.606 \\
 & (0.429) & (0.421) & (0.393) & (0.388) \\
 YCuO$_{3-x}$ & 0.399  & 0.386 & 0.355 & 0.350 \\
 & (0.687) & (0.664) & (0.611) & (0.602) \\
LaCuO$_{3-x}$ & 0.283  & 0.274 & 0.255 & 0.250 \\
 & (1.033) & (1.000) & (0.932) & (0.912) \\
\hline
$N_c$ & 13 & 21 & 36 & 52 \\
\end{tabular}
\end{ruledtabular}
\end{table}

\begin{figure}[tbp]
\includegraphics[clip,width=0.7\linewidth]{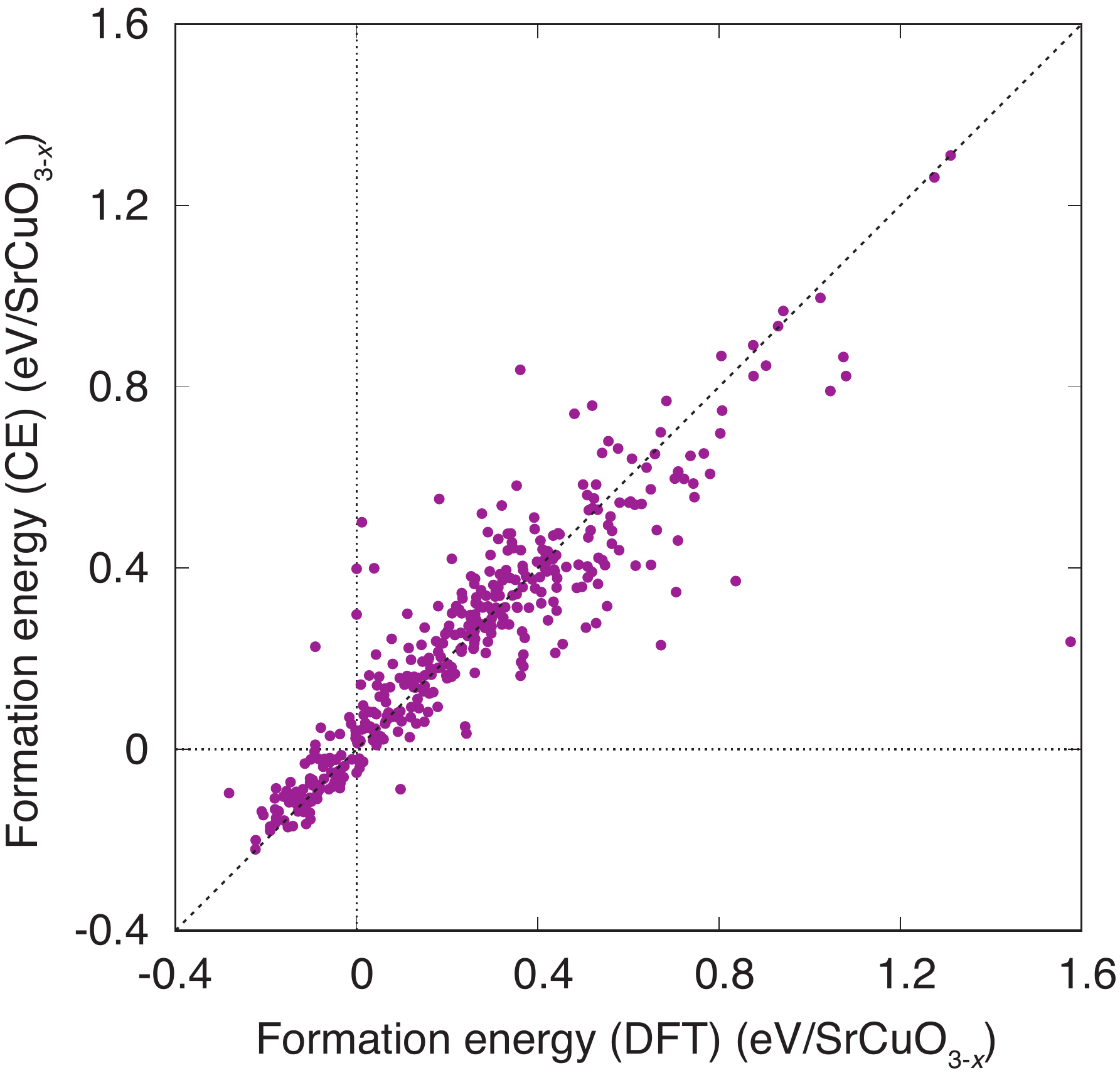}
\caption{
Distribution of energies of nonequivalent oxygen-deficient structures with up to $n=4$ predicted using a CE model with 36 clusters up to fourbodies (model 3 in Table \ref{perovskite:error-ce}) in SrCuO$_{3-x}$.
The Pearson correlation coefficient between the DFT and CE energies is $R=0.90$.
}
\label{perovskite:energy-distribution-CE}
\end{figure}

Then, we construct various CE models for each system using the training dataset.
We employ spin values of $+1$ and $-1$ for the oxygen atom and vacancy, respectively, which are required to define the correlation function.
Table \ref{perovskite:error-ce} shows the fitting errors of the four CE models, which indicates that all of the CE models have significant errors.
They are much more significant than the order of the difference of the formation energies between structures.
Figure \ref{perovskite:energy-distribution-CE} also shows the distribution of the formation energies of nonequivalent oxygen-deficient structures estimated using a CE model for SrCuO$_{3-x}$.
The formation energies deviate from those estimated by DFT calculations.
Therefore, it is impossible to find the true ground states using only the energy predicted using the CE model.

The significant error in the CE model arises from the local geometry optimization of the oxygen-deficient structures in the DFT calculation. 
Although the local geometry optimization is essential for predicting oxygen-deficient structures, the CE method describes the energy of the optimized structure using a linear polynomial model of the correlation functions defined for the ideal lattice of cubic perovskite.
A general procedure of the CE method has been successfully applied to most of the substitutional alloys, while it does not work well in systems involving the significant atomic relaxation from the ideal lattice sites in a straightforward manner.

\section{Bayesian optimization}
\label{perovskite:section-bo}
\subsection{Gaussian process regression}

To find stable oxygen-deficient perovskite structures, we employ Bayesian optimization \cite{jones01} based on GP \cite{Rasmussen_2006} specified by its mean function and covariance function instead of a linear CE model.
Representing a structure by vector $\bm{d}$, a radial basis function covariance or kernel between structures $\bm{d}_i$ and $\bm{d}_j$ for noise-free observation is given by
\begin{equation}
k\left(\bm{d}_i, \bm{d}_j \right) = \sigma_f^2  \exp \left( - \frac{|\bm{d}_i - \bm{d}_j|^2}{2 l^2} \right), 
\end{equation}
where $l$ and $\sigma_f^2$ are tuning parameters controlling the length scales for structure $\bm{d}$ and the observation, respectively.
Here, we represent oxygen-deficient structures by the correlation functions of point clusters and 11 pair clusters used in the CE models shown in Fig. \ref{perovskite:structure-correlation} (c).
As described above, the discrepancy between the DFT calculation and the CE model is too large to predict the stable oxygen-deficient perovskite structures using the CE model itself.
On the other hand, the CE models have a strong correlation with the DFT calculation, as shown in Fig. \ref{perovskite:energy-distribution-CE}.
This implies that a set of correlation functions is useful for representing a structure in GP-based Bayesian optimization.

Describing the formation energies of structures in the training dataset as $\bm{y}$, the mean function $\mu$ at structure $\bm{d}_*$ and the variance function $\sigma_*^2$ are given as
\begin{equation}
\mu(\bm{d}_*) = {\bm k}_*^\top \bm{K}^{-1} \bm{y}
\end{equation}
and
\begin{equation}
\sigma_*^2 = k(\bm{d}_*,\bm{d}_*) - {\bm k}_*^\top \bm{K}^{-1} {\bm k}_*,
\end{equation}
respectively, where ${\bm k}_* = \left[ k(\bm{d}_*, \bm{d}_1), \cdots, k(\bm{d}_*, \bm{d}_N) \right] ^\top$ is the vector of kernel functions between structure $\bm{d}_*$ and $N$ structures in the training dataset.
A symmetric kernel matrix $\bm{K}$ is composed of kernel functions for all pair arrangements of the training data, expressed as
\begin{equation}
\bm{K} =
\begin{pmatrix}
k\left(\bm{d}_1, \bm{d}_1 \right) & k\left(\bm{d}_1, \bm{d}_2 \right) & \cdots & k\left(\bm{d}_1, \bm{d}_N \right) \\
k\left(\bm{d}_2, \bm{d}_1 \right) & k\left(\bm{d}_2, \bm{d}_2 \right) & \cdots & k\left(\bm{d}_2, \bm{d}_N \right) \\
\vdots & \vdots & \ddots & \vdots \\
k\left(\bm{d}_N, \bm{d}_1 \right) & k\left(\bm{d}_N, \bm{d}_2 \right) & \cdots & k\left(\bm{d}_N, \bm{d}_N \right) \\
\end{pmatrix}
.
\end{equation}

\subsection{Procedure of Bayesian optimization}
Our procedure of Bayesian optimization is as follows.
First, a GP model is developed from an initial set of formation energies. 
The initial set is composed of the formation energies of all nonequivalent structures with up to $n=4$ and 100 nonequivalent structures that are randomly selected from the pool of 773,613 nonequivalent structures with up to $n=8$.
The model is then iteratively updated by (i) sampling the structure for which the formation energy is expected to be the lowest among the remaining candidate structures for each composition and (ii) updating the GP model using the training dataset updated by the structures observed in step (i).
These steps are repeated until the convex hull of the formation energy converges.
We exclude some structures from the dataset, i.e., structures for which DFT calculations fail to converge.

There are several well-known procedures to sample the structure that is expected to show the lowest formation energy in step (i).
Here, we adopt a simple procedure of the probability of improvement (PI) strategy \cite{jones01}.
In the minimization problem of formation energy, the PI involves sampling the structure at which the probability that the formation energy is lower than $y_{\rm best}$ is maximized, where $y_{\rm best}$ denotes the current best formation energy among the training data.
This means that structure $\bm{d}_{i'}$ is selected by maximizing the probability as 
\begin{equation}
\bm{d}_{i'}: = \mathop{\rm argmax}\limits_{\bm{d}_*} \Phi \left(\frac{y_{\rm best}-\mu(\bm{d}_*)}{\sigma_*} \right),
\end{equation}
where $\Phi \left(y-\mu(\bm{d}_*)/\sigma_* \right)$ denotes the cumulative distribution function of normal distribution $N(\mu, \sigma^2)$.

\subsection{Oxygen-deficient perovskite structures at 0 GPa}
\label{perovskite:sec-0GPa}

\begin{figure}[tbp]
\includegraphics[clip,width=\linewidth]{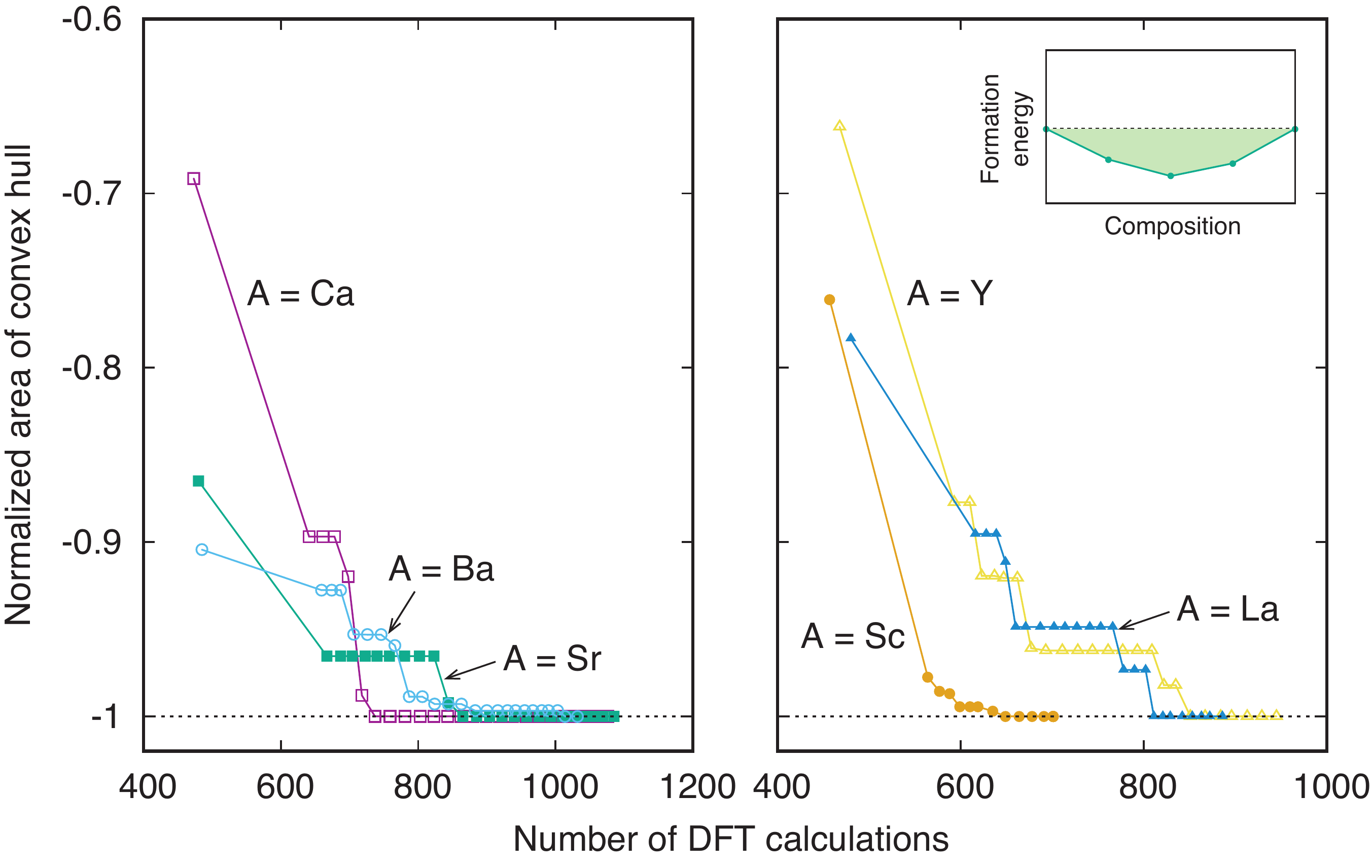}
\caption{
Convergence behavior of the convex hull of formation energy in Bayesian optimization for ACuO$_{3-x}$ systems.
The vertical axis indicates the area of the convex hull below the line indicating zero formation energy, schematically illustrated as the shaded area in the top of the right panel.
The area is normalized by the area of the converged convex hull.
}
\label{perovskite:history-bo}
\end{figure}

\begin{figure*}[tbp]
\includegraphics[clip,width=0.8\linewidth]{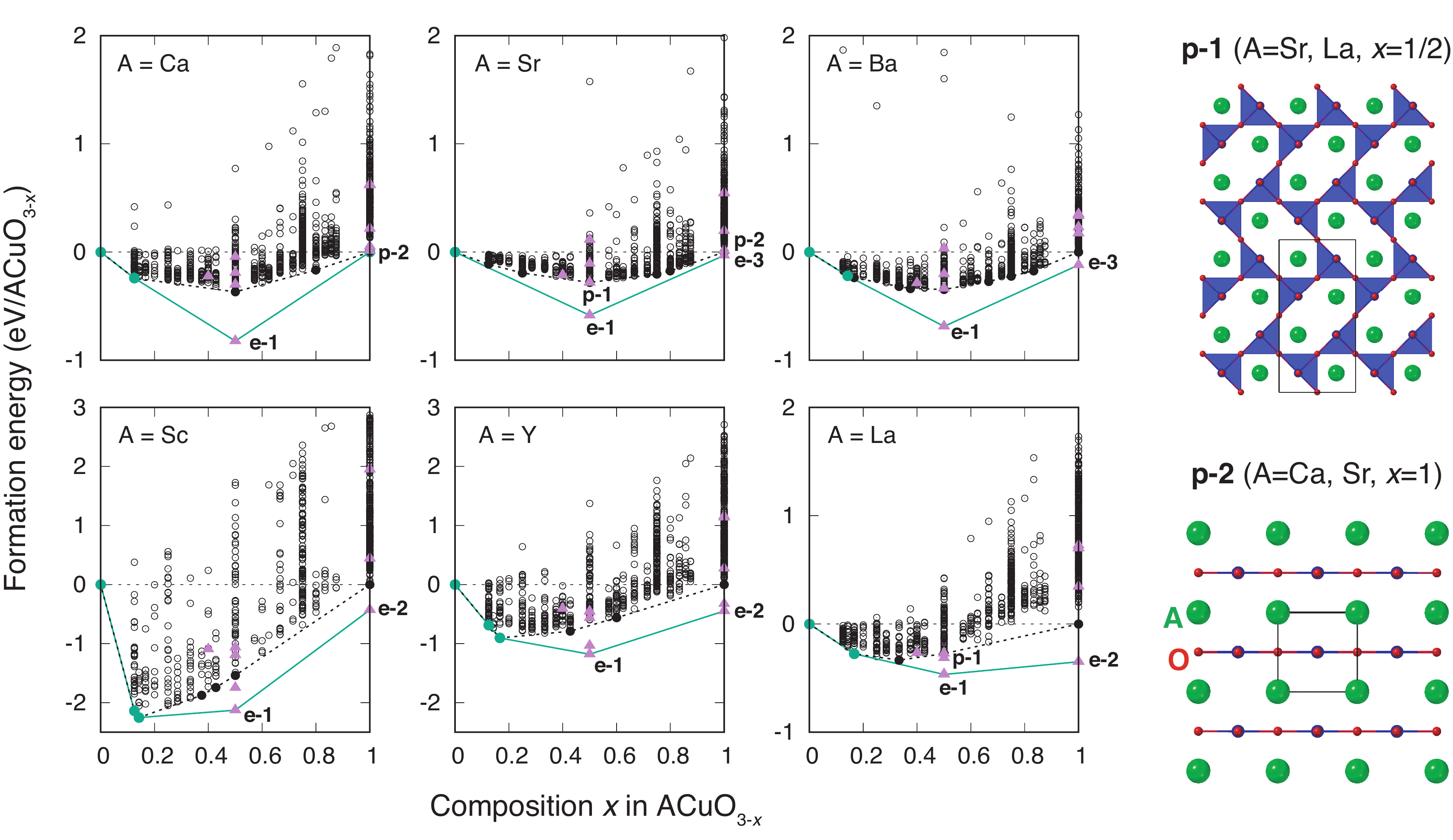}
\caption{
Formation energy distribution of oxygen-deficient perovskite structures with up to $n=8$ selected by Bayesian optimization.
The broken black line and black closed circles show the convex hull of formation energy and its vertices, respectively, obtained only from oxygen-deficient perovskite structures.
The purple closed triangles show the formation energies of experimental prototype structures.
The green line shows the convex hull obtained from the combination of oxygen-deficient perovskite structures and the experimental prototype structures.
The oxygen-deficient perovskite structures reported in the literature on the convex hulls are indicated by \emph{p-1} (Ca$_2$Mn$_2$O$_5$ type) and \emph{p-2} (LaNiO$_{2}$ type). 
Their crystal structures are also shown. 
Moreover, the experimental structures on the convex hulls are indicated as \emph{e-1} (Sc$_2$Cu$_2$O$_5$ type), \emph{e-2} (delafosite FeCuO$_2$ type), and \emph{e-3} (SrCuO$_2$ type).
}
\label{perovskite:energy-8}
\end{figure*}

Figure \ref{perovskite:history-bo} shows the convergence behavior of the convex hull of formation energy in Bayesian optimization.
We measure the convergence degree of the convex hull using the area of the convex hull below the line indicating zero formation energy.
As can be seen in Fig. \ref{perovskite:history-bo}, the convex hull converges well in all systems.
The numbers of DFT calculations required for the convergence are 1076, 1085, 1032, 701, 945, and 886 in CaCuO$_{3-x}$, SrCuO$_{3-x}$, BaCuO$_{3-x}$, ScCuO$_{3-x}$, YCuO$_{3-x}$, and LaCuO$_{3-x}$, respectively.

Figure \ref{perovskite:energy-8} shows the formation energies of the oxygen-deficient structures with up to $n=8$ selected by Bayesian optimization including the structures in the initial training dataset of Bayesian optimization.
The broken line in Fig. \ref{perovskite:energy-8} indicates the convex hull of formation energy representing the set of oxygen-deficient perovskite structures that are more stable than the other oxygen-deficient perovskite structures.
We have found three, five, eight, five, four and three oxygen-deficient perovskite structures on the convex hull in CaCuO$_{3-x}$, SrCuO$_{3-x}$, BaCuO$_{3-x}$, ScCuO$_{3-x}$, YCuO$_{3-x}$, and LaCuO$_{3-x}$, respectively.
Four of the oxygen-deficient perovskite structures on the convex hull, i.e., CaCuO$_{2}$, SrCuO$_{2}$, SrCuO$_{2.5}$, and LaCuO$_{2.5}$, are identical to the experimental structures in the literature \cite{CHEN1996498,bringley1990synthesis,KARPINSKI199410,TAKANO1989375}.
They are denoted by \emph{p-1} and \emph{p-2} in Fig. \ref{perovskite:energy-8}.

\subsection{Stability relative to experimental structures}
Although the convex hull of oxygen-deficient perovskite structures has many vertices or oxygen-deficient perovskite structures, experimental structures that are not classified into the perovskite family have also been known in ACuO$_{3-x}$ systems.
In the six ACuO$_{3-x}$ systems, seven structure prototypes with no partial occupancy of Sc$_2$Cu$_2$O$_5$, Y$_2$Cu$_2$O$_5$, Nd$_2$Cu$_2$O$_5$, SrCuO$_2$, BaCuO$_2$, delafosite FeCuO$_2$, and CaCuO$_2$ types are known in addition to the oxygen-deficient perovskite Ca$_2$Mn$_2$O$_5$ type (\emph{p-1}) and LaNiO$_2$ type (\emph{p-2}).
Therefore, we estimate the formation energy for the seven prototypes in each system by DFT calculation.

Figure \ref{perovskite:energy-8} shows the formation energy distribution of the experimental prototype structures.
The convex hull obtained from the oxygen-deficient perovskite structures and the experimental prototype structures is also shown in Fig. \ref{perovskite:energy-8}.
In all compositions of ACuO$_{2.5}$, the Sc$_2$Cu$_2$O$_5$ type structure (\emph{e-1}) is stable although the Y$_2$Cu$_2$O$_5$-type structure is competitive with the Sc$_2$Cu$_2$O$_5$-type structure in LaCuO$_{2.5}$.
Also, the SrCuO$_2$ type (\emph{e-3}) is stable in SrCuO$_2$ and BaCuO$_2$, whereas the delafosite FeCuO$_2$ type (\emph{e-2}) is stable in ScCuO$_2$, YCuO$_2$, and LaCuO$_2$.
The stable structure of CaCuO$_2$ is of the LaNiO$_2$ type (\emph{p-2}), which is consistent with the experimental structure.
On the other hand, most of the oxygen-deficient perovskite structures are not vertices of the convex hull, which indicates that they are not thermodynamically stable at 0 GPa.

\subsection{Clustering of oxygen-deficient perovskite structures}

\begin{figure}[tbp]
\includegraphics[clip,width=0.95\linewidth]{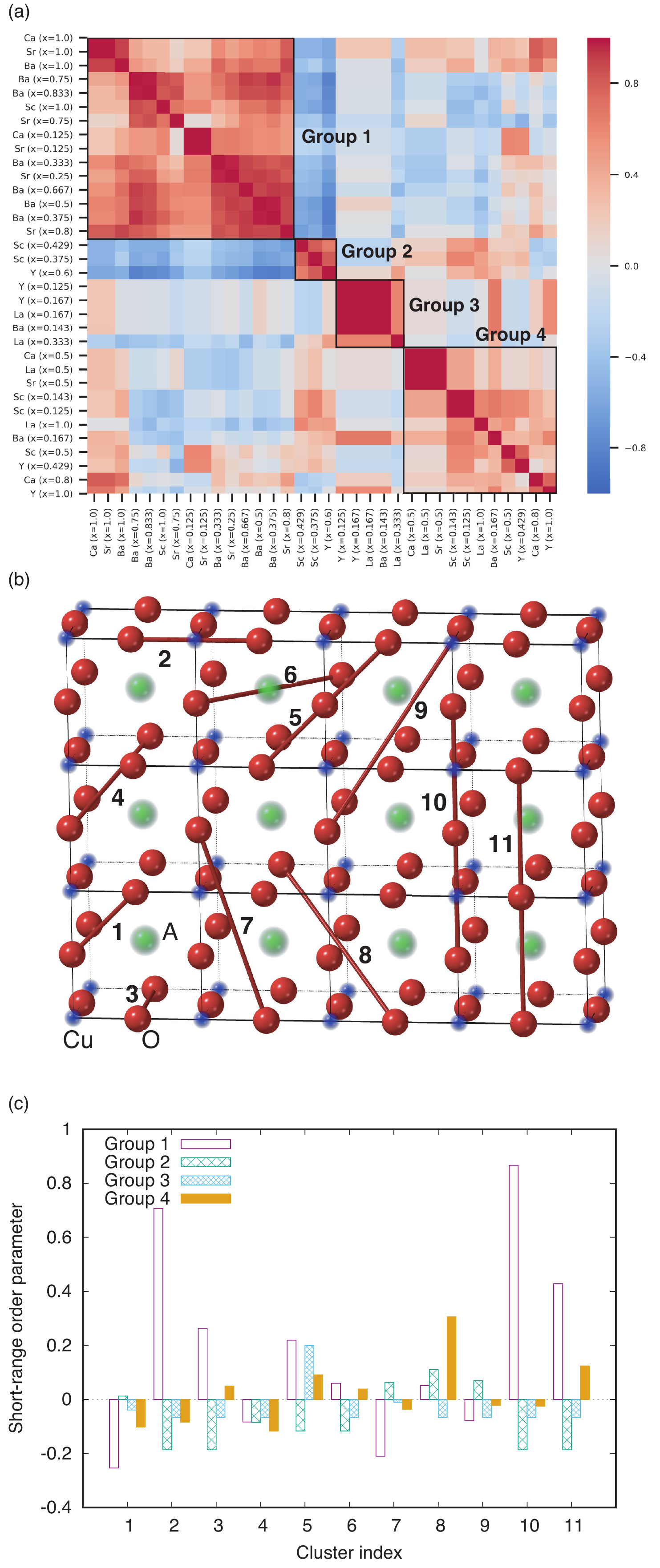}
\caption{
(a) Pearson correlation coefficients between oxygen-deficient perovskite structures on the convex hull.
(b) Pair clusters used in the CE method and Bayesian optimization. 
(c) Dependence of WC-SRO parameters on the group classified by a hierarchical clustering.
A bar shows the WC-SRO parameter averaged over the oxygen-deficient perovskite structures belonging to each group.
The cluster indexes correspond to those of pairs shown in (b).
}
\label{perovskite:structure-correlation}
\end{figure}

To understand the common features of the oxygen-deficient perovskite structures on the convex hull, we carry out a machine learning clustering of the oxygen-deficient perovskite structures.
Here, we introduce the Warren--Cowley short-range order (WC-SRO) parameter to represent the oxygen-deficient perovskite structures.
The WC-SRO parameters can be derived from the correlation functions used in both the CE method and Bayesian optimization.
In ACuO$_{3-x}$, the WC-SRO parameter of pair $\alpha$ for structure $\bm{d}$ is defined as
\begin{equation}
{\rm SRO}_\alpha (\bm{d}) = \frac{\varphi_\alpha(\bm{d})-q^2}{1-q^2},
\end{equation}
where $\varphi_\alpha(\bm{d})$ denotes the correlation function of pair $\alpha$ and $q=(3-2x)/3$.

We employ a hierarchical clustering approach using Pearson correlation coefficients between the oxygen-deficient perovskite structures.
For the distance metric to define the similarity between the structures, we adopt a complete-linkage clustering, which uses the maximum correlation-based distances between all observations of the two sets.
Figure \ref{perovskite:structure-correlation} (a) shows Pearson correlation coefficients between the oxygen-deficient perovskite structures on the convex hull, represented by WC-SRO parameters of the eleven pairs shown in Fig. \ref{perovskite:structure-correlation} (b).
As can be seen in Fig. \ref{perovskite:structure-correlation} (a), the hierarchical clustering classifies the oxygen-deficient perovskite structures into four large groups.

Figure \ref{perovskite:structure-correlation} (c) shows the WC-SRO parameters averaged over the oxygen-deficient perovskite structures belonging to each group.
Group 1 shows positive large values of the WC-SRO parameters for the second, tenth, and eleventh pairs; hence, most of those pairs are composed of the same species in the structures of group 1. 
This indicates that the structures in group 1 consist mainly of planar fourfold copper units.
An example is SrCuO$_2$ with the \emph{p-2} structure illustrated in Fig. \ref{perovskite:energy-8}, where all copper atoms are planar fourfold units.
Also, group 4 is characterized by a positive value of the WC-SRO parameter of the nearest pair along the $\langle111\rangle$ direction (the eighth pair).
This means that $\langle111\rangle$ nearest pairs of vacancies are components of the structures in group 4.
An example is SrCuO$_{2.5}$ with the \emph{p-1} structure illustrated in Fig. \ref{perovskite:energy-8}.

Note that we evaluate WC-SRO parameters of the oxygen-deficient perovskite structures with the ideal cubic perovskite lattice.
Some of the structures optimized by the DFT calculation significantly differ from their ideal structures.

\begin{figure*}[tbp]
\includegraphics[clip,width=0.6\linewidth]{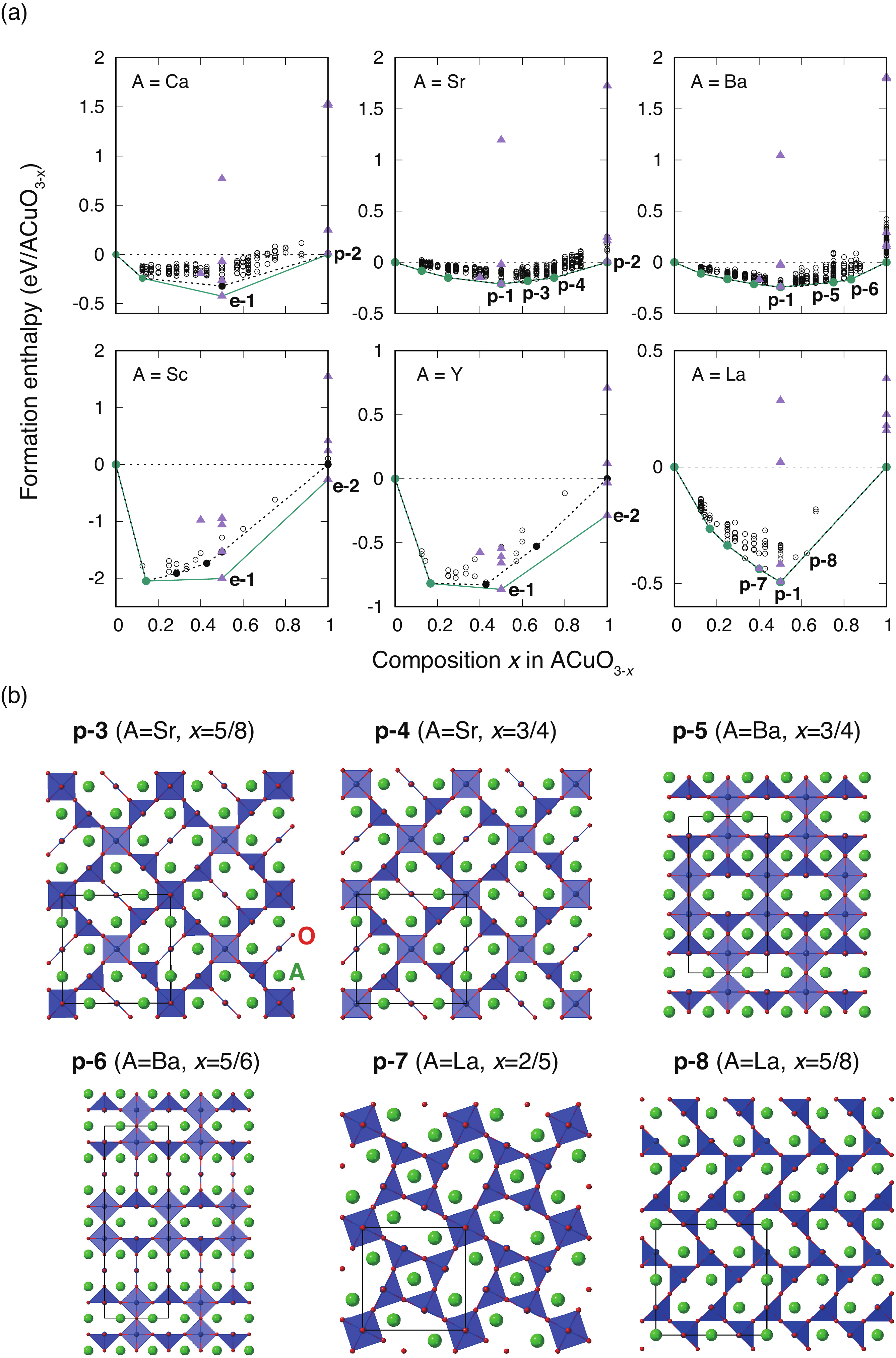}
\caption{
(a) Formation enthalpy distribution at 10 GPa of the oxygen-deficient perovskite structures satisfying the condition that their energies measured from the convex hull of the oxygen-deficient perovskite structures are less than 0.1 eV/ACuO$_{3-x}$.
The broken black line and closed circles show the convex hull of the oxygen-deficient perovskite structures and its vertices, respectively.
The purple closed triangles show the formation enthalpy values of experimental prototype structures.
The green line shows the convex hull obtained from the combination of the oxygen-deficient perovskite structures and the experimental prototype structures.
(b) Crystal structures of the oxygen-deficient perovskites with composition $x \geq 0.4$ on the convex hull or almost on the convex hull.
Only LaCuO$_{2.6}$ (\emph{p-7}) is experimentally reported \cite{bringley1990synthesis}. 
It has the crystal structure type identical to that of SrMnO$_{2.6}$ \cite{SUESCUN20071698}.
}
\label{perovskite:energy-10GPa}
\end{figure*}

\subsection{Stability at 10 GPa}

As shown in Sec. \ref{perovskite:sec-0GPa}, most of the oxygen-deficient perovskite structures are not stable at 0 GPa.
On the other hand, oxygen-deficient perovskite compounds such as CaCuO$_2$, SrCuO$_2$, and SrCuO$_{2.5}$ have been synthesized under high-pressure conditions \cite{CHEN1996498,KARPINSKI199410,TAKANO1989375}.
Therefore, we investigate the pressure dependence of the phase stability of the oxygen-deficient perovskite structures.
We restrict the DFT calculation to the experimental prototype structures and the oxygen-deficient perovskite structures satisfying the condition that their energies measured from the convex hull of the oxygen-deficient perovskite structures are less than 0.1 eV/ACuO$_{3-x}$.

Figure \ref{perovskite:energy-10GPa} (a) shows the formation enthalpy distribution of the experimental structures and the oxygen-deficient perovskite structures at 10 GPa.
Contrary to the phase stability at 0 GPa, all of the experimental oxygen-deficient perovskite structures, namely, CaCuO$_2$, SrCuO$_2$, SrCuO$_{2.5}$, LaCuO$_{2.5}$, and LaCuO$_{2.6}$, are stable at 10 GPa.
In addition, the convex hull indicates that some of the oxygen-deficient perovskite structures are stable at 10 GPa in SrCuO$_{3-x}$, BaCuO$_{3-x}$, and LaCuO$_{3-x}$. 
Figure \ref{perovskite:energy-10GPa} (b) shows the crystal structures of the oxygen-deficient perovskites with the composition $x \geq 0.4$ on or almost on the convex hull, where only LaCuO$_{2.6}$ (\emph{p-7}) is experimentally reported \cite{bringley1990synthesis}.
The crystal structures with the other compositions are also shown in Appendix.
In CaCuO$_{3-x}$, ScCuO$_{3-x}$, and YCuO$_{3-x}$, the structures on the convex hull at 10 GPa are almost the same as those at 0 GPa.
However, some oxygen-deficient perovskite structures such as CaCuO$_{2.5}$ and YCuO$_{2.571}$ are close to the convex hull.
They are expected to be stable at higher pressures.

As can be seen in Fig. \ref{perovskite:energy-10GPa}, the oxygen configuration in the predicted structure of SrCuO$_{2.375}$ (\emph{p-3}) is similar to that in the experimental structure of La$_{0.5}$Ca$_{0.5}$CuO$_{2.5}$ \cite{GULOY199454}. 
The difference between them is the Cu--O polyhedron located at the center of the unit cell, i.e., they are planar fourfold and octahedral sixfold structures, respectively. 
The predicted structure of SrCuO$_{2.25}$ (\emph{p-4}) is also similar to the structures of SrCuO$_{2.375}$ (\emph{p-3}) and La$_{0.5}$Ca$_{0.5}$CuO$_{2.5}$. 
At the same time, since La$_{0.5}$Ca$_{0.5}$CuO$_{2.25}$ \cite{GULOY199454} and La$_{0.25}$Sr$_{0.75}$CuO$_{2.25}$ \cite{FU1990291} are reported as existing compounds, it seems to be natural to compare SrCuO$_{2.25}$ (\emph{p-4}) with them.
However, their crystal structures remain unclarified.
The results in Ref. \onlinecite{FU1990291} only reveal that La$_{0.25}$Sr$_{0.75}$CuO$_{2.25}$ has the space group of $P$42$_1$2 or $P\bar 4$2$_1 m$. 
Nevertheless, they are subgroups of the space group $P$4/$mbm$ of SrCuO$_{2.25}$ (\emph{p-4}) with index 2. 
Hence, the oxygen configuration in SrCuO$_{2.25}$ (\emph{p-4}) may be identical to those in La$_{0.5}$Ca$_{0.5}$CuO$_{2.25}$ and La$_{0.25}$Sr$_{0.75}$CuO$_{2.25}$.

Besides, we compare the oxygen-deficient structures of BaCuO$_{2.333}$ with that of YBa$_2$Cu$_3$O$_{7-d}$ (YBCO), because the present structure dataset contains the oxygen configuration of YBCO.
As shown in Appendix (Fig. \ref{perovskite:structure-convex1}), although the YBCO-like structure cannot be found as an oxygen-deficient structure with the lowest enthalpy at $x=2/3$ for BaCuO$_{3-x}$, the formation enthalpy of the YBCO-like structure is only 52 meV/BaCuO$_{3-x}$ larger than that of the lowest formation enthalpy. 
This implies that the formation of the YBCO-type structure consisting of the alternate stacking of Cu--O chains and Cu--O double layers is dominated not by the cation ordering but by the oxygen vacancy ordering.

\subsection{Pressure dependence of stability}

\begin{figure}[tbp]
\includegraphics[clip,width=0.7\linewidth]{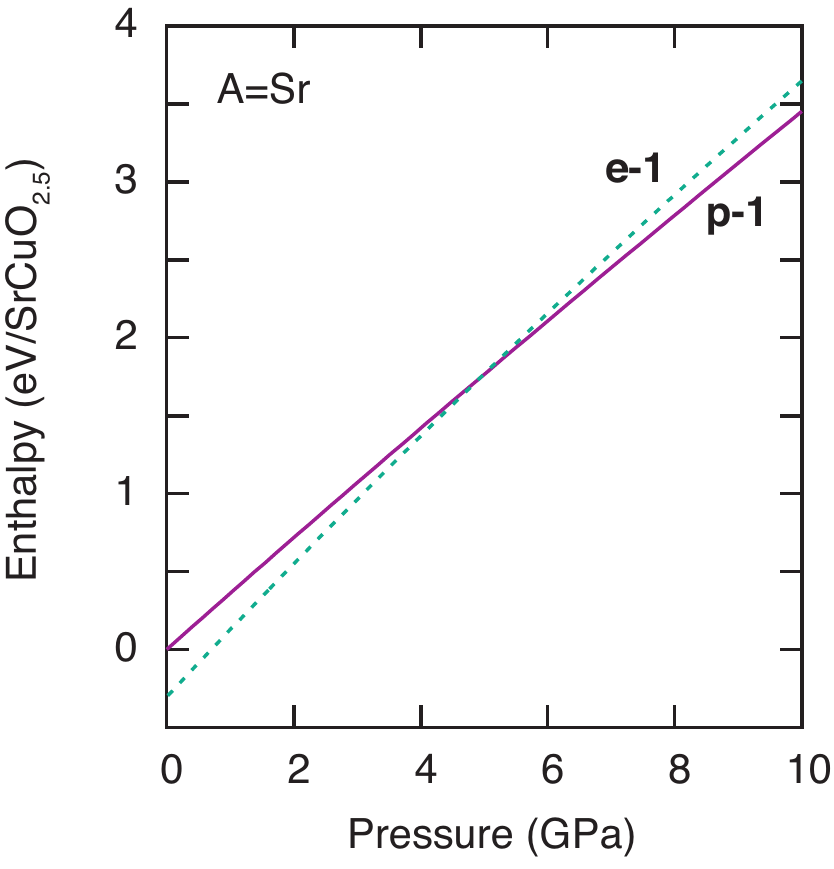}
\caption{
Pressure--enthalpy curves of the oxygen-deficient perovskite structure (\emph{p-1}) and Sc$_2$Cu$_2$O$_5$-type structure (\emph{e-1}) in SrCuO$_{2.5}$.
}
\label{perovskite:enthalpy}
\end{figure}

Finally, we demonstrate the pressure dependence of phase stability between the oxygen-deficient perovskite structures and the experimental structures.
For each structure, the pressure--enthalpy curve is obtained from the energy--volume curve fitted to the Vinet equation of states \cite{doi:10.1029/JB092iB09p09319} using the DFT energies of ten structures constructed by expansions/contractions of the equilibrium structure.
Figure \ref{perovskite:enthalpy} shows the pressure--enthalpy curves of the oxygen-deficient perovskite structure (\emph{p-1}) and Sc$_2$Cu$_2$O$_5$-type structure (\emph{e-1}) in SrCuO$_{2.5}$.
As can be seen in Fig. \ref{perovskite:enthalpy}, the oxygen-deficient perovskite structure becomes stable as the pressure increases.
It is found that the transition pressure from the Sc$_2$Cu$_2$O$_5$-type structure to the oxygen-deficient perovskite \emph{p-1} structure is 5.6 GPa in SrCuO$_{2.5}$.
Similarly, LaCuO$_{2.5}$ involves the phase transition from the Sc$_2$Cu$_2$O$_5$-type structure to the Nd$_2$Cu$_2$O$_5$-type structure at 1.9 GPa and the phase transition from the Nd$_2$Cu$_2$O$_5$-type structure to the oxygen-deficient perovskite \emph{p-1} structure at 4.4 GPa.

\section{Conclusion}
The stability of oxygen-deficient perovskite structures in ACuO$_{3-x}$ (A $=$ Ca, Sr, Ba, Sc, Y, La) has been investigated systematically using a combination of DFT calculation, the CE method, GP, and Bayesian optimization.
We have successfully reproduced the reported oxygen-deficient perovskite structures and found a series of stable oxygen-deficient structures that are not reported in the literature. 
Our results indicate that high-pressure synthesis is advantageous for obtaining the oxygen-deficient perovskite structures. 
By adopting the machine learning clustering for the obtained structures, we found that they can be classified into four groups and one of the largest groups shows a common structural feature characterized by the square planar coordination for copper.

Regarding the computational aspects of structure prediction, the CE method fails to derive an accurate model for the formation energy of the oxygen-deficient structure owing to its substantial geometry relaxation from the ideal cubic perovskite lattice.
On the other hand, the combination of the CE method, GP model, and Bayesian optimization is very useful for predicting the oxygen-deficient perovskite structures.
This procedure is also applicable in a straightforward manner to a wide range of atom-deficient structures involving extensive geometry relaxation from their ideal lattice.

\begin{acknowledgments}
This work was supported by a Grant-in-Aid for Scientific Research (B) (Grant Number 19H02419), a Grant-in-Aid for Challenging Research (Exploratory) (Grant Number 18K18942), and a Grant-in-Aid for Scientific Research on Innovative Areas (Grant Number 19H05787) from the Japan Society for the Promotion of Science (JSPS).
\end{acknowledgments}

\appendix
\section{Hermite normal forms}

Table \ref{perovskite:list-hnf} shows a complete set of nonequivalent HNFs with up to $n=8$ for the cubic perovskite lattice.

\begin{table*}[tbp]
\begin{ruledtabular}
\caption{
Complete set of nonequivalent HNFs for the cubic perovskite lattice.
}
\label{perovskite:list-hnf}
\begin{tabular}{c|l}
Determinant of HNF & Nonequivalent HNF \\
\hline
$n=1$ & $
\begin{pmatrix}
1 & 0 & 0 \\
0 & 1 & 0 \\
0 & 0 & 1 \\
\end{pmatrix}
$ \\

\hline
$n=2$ & $
\begin{pmatrix}
1 & 0 & 0 \\
0 & 1 & 0 \\
0 & 0 & 2 \\
\end{pmatrix}
\begin{pmatrix}
1 & 0 & 0 \\
0 & 1 & 0 \\
0 & 1 & 2 \\
\end{pmatrix}
\begin{pmatrix}
1 & 0 & 0 \\
0 & 1 & 0 \\
1 & 1 & 2 \\
\end{pmatrix}
$ \\

\hline
$n=3$ & $
\begin{pmatrix}
1 & 0 & 0 \\
0 & 1 & 0 \\
0 & 0 & 3 \\
\end{pmatrix}
\begin{pmatrix}
1 & 0 & 0 \\
0 & 1 & 0 \\
0 & 1 & 3 \\
\end{pmatrix}
\begin{pmatrix}
1 & 0 & 0 \\
0 & 1 & 0 \\
1 & 1 & 3 \\
\end{pmatrix}
$ \\

\hline
$n=4$ & $
\begin{pmatrix}
1 & 0 & 0 \\
0 & 1 & 0 \\
0 & 0 & 4 \\
\end{pmatrix}
\begin{pmatrix}
1 & 0 & 0 \\
0 & 1 & 0 \\
0 & 1 & 4 \\
\end{pmatrix}
\begin{pmatrix}
1 & 0 & 0 \\
0 & 1 & 0 \\
0 & 2 & 4 \\
\end{pmatrix}
\begin{pmatrix}
1 & 0 & 0 \\
0 & 1 & 0 \\
1 & 1 & 4 \\
\end{pmatrix}
\begin{pmatrix}
1 & 0 & 0 \\
0 & 1 & 0 \\
1 & 2 & 4 \\
\end{pmatrix}
\begin{pmatrix}
1 & 0 & 0 \\
0 & 1 & 0 \\
2 & 2 & 4 \\
\end{pmatrix}
\begin{pmatrix}
1 & 0 & 0 \\
0 & 2 & 0 \\
0 & 0 & 2 \\
\end{pmatrix}
\begin{pmatrix}
1 & 0 & 0 \\
0 & 2 & 0 \\
1 & 0 & 2 \\
\end{pmatrix}
\begin{pmatrix}
1 & 0 & 0 \\
1 & 2 & 0 \\
1 & 0 & 2 \\
\end{pmatrix}
$ \\

\hline
$n=5$ & $
\begin{pmatrix}
1 & 0 & 0 \\
0 & 1 & 0 \\
0 & 0 & 5 \\
\end{pmatrix}
\begin{pmatrix}
1 & 0 & 0 \\
0 & 1 & 0 \\
0 & 1 & 5 \\
\end{pmatrix}
\begin{pmatrix}
1 & 0 & 0 \\
0 & 1 & 0 \\
0 & 2 & 5 \\
\end{pmatrix}
\begin{pmatrix}
1 & 0 & 0 \\
0 & 1 & 0 \\
1 & 1 & 5 \\
\end{pmatrix}
\begin{pmatrix}
1 & 0 & 0 \\
0 & 1 & 0 \\
1 & 2 & 5 \\
\end{pmatrix}
$ \\

\hline
$n=6$ & 
\begin{tabular}{l}
$
\begin{pmatrix}
1 & 0 & 0 \\
0 & 1 & 0 \\
0 & 0 & 6 \\
\end{pmatrix}
\begin{pmatrix}
1 & 0 & 0 \\
0 & 1 & 0 \\
0 & 1 & 6 \\
\end{pmatrix}
\begin{pmatrix}
1 & 0 & 0 \\
0 & 1 & 0 \\
0 & 2 & 6 \\
\end{pmatrix}
\begin{pmatrix}
1 & 0 & 0 \\
0 & 1 & 0 \\
0 & 3 & 6 \\
\end{pmatrix}
\begin{pmatrix}
1 & 0 & 0 \\
0 & 1 & 0 \\
1 & 1 & 6 \\
\end{pmatrix}
\begin{pmatrix}
1 & 0 & 0 \\
0 & 1 & 0 \\
1 & 2 & 6 \\
\end{pmatrix}
\begin{pmatrix}
1 & 0 & 0 \\
0 & 1 & 0 \\
1 & 3 & 6 \\
\end{pmatrix}
\begin{pmatrix}
1 & 0 & 0 \\
0 & 1 & 0 \\
2 & 2 & 6 \\
\end{pmatrix}
\begin{pmatrix}
1 & 0 & 0 \\
0 & 1 & 0 \\
2 & 3 & 6 \\
\end{pmatrix}$
\\
$\begin{pmatrix}
1 & 0 & 0 \\
0 & 1 & 0 \\
3 & 3 & 6 \\
\end{pmatrix}
\begin{pmatrix}
1 & 0 & 0 \\
0 & 2 & 0 \\
0 & 0 & 3 \\
\end{pmatrix} 
\begin{pmatrix}
1 & 0 & 0 \\
0 & 2 & 0 \\
1 & 0 & 3 \\
\end{pmatrix}
\begin{pmatrix}
1 & 0 & 0 \\
1 & 2 & 0 \\
0 & 0 & 3 \\
\end{pmatrix}
$ 
\end{tabular} \\

\hline
$n=7$ & 
\begin{tabular}{l}
$
\begin{pmatrix}
1 & 0 & 0 \\
0 & 1 & 0 \\
0 & 0 & 7 \\
\end{pmatrix}
\begin{pmatrix}
1 & 0 & 0 \\
0 & 1 & 0 \\
0 & 1 & 7 \\
\end{pmatrix}
\begin{pmatrix}
1 & 0 & 0 \\
0 & 1 & 0 \\
0 & 2 & 7 \\
\end{pmatrix}
\begin{pmatrix}
1 & 0 & 0 \\
0 & 1 & 0 \\
1 & 1 & 7 \\
\end{pmatrix}
\begin{pmatrix}
1 & 0 & 0 \\
0 & 1 & 0 \\
1 & 2 & 7 \\
\end{pmatrix}
\begin{pmatrix}
1 & 0 & 0 \\
0 & 1 & 0 \\
1 & 3 & 7 \\
\end{pmatrix}
\begin{pmatrix}
1 & 0 & 0 \\
0 & 1 & 0 \\
2 & 3 & 7 \\
\end{pmatrix}
$ 
\end{tabular} \\

\hline
$n=8$ & 
\begin{tabular}{l}
$
\begin{pmatrix}
1 & 0 & 0 \\
0 & 1 & 0 \\
0 & 0 & 8 \\
\end{pmatrix}
\begin{pmatrix}
1 & 0 & 0 \\
0 & 1 & 0 \\
0 & 1 & 8 \\
\end{pmatrix}
\begin{pmatrix}
1 & 0 & 0 \\
0 & 1 & 0 \\
0 & 2 & 8 \\
\end{pmatrix}
\begin{pmatrix}
1 & 0 & 0 \\
0 & 1 & 0 \\
0 & 3 & 8 \\
\end{pmatrix}
\begin{pmatrix}
1 & 0 & 0 \\
0 & 1 & 0 \\
0 & 4 & 8 \\
\end{pmatrix}
\begin{pmatrix}
1 & 0 & 0 \\
0 & 1 & 0 \\
1 & 1 & 8 \\
\end{pmatrix}
\begin{pmatrix}
1 & 0 & 0 \\
0 & 1 & 0 \\
1 & 2 & 8 \\
\end{pmatrix}
\begin{pmatrix}
1 & 0 & 0 \\
0 & 1 & 0 \\
1 & 3 & 8 \\
\end{pmatrix}
\begin{pmatrix}
1 & 0 & 0 \\
0 & 1 & 0 \\
1 & 4 & 8 \\
\end{pmatrix}$
\\
$\begin{pmatrix}
1 & 0 & 0 \\
0 & 1 & 0 \\
2 & 2 & 8 \\
\end{pmatrix}
\begin{pmatrix}
1 & 0 & 0 \\
0 & 1 & 0 \\
2 & 3 & 8 \\
\end{pmatrix} 
\begin{pmatrix}
1 & 0 & 0 \\
0 & 1 & 0 \\
2 & 4 & 8 \\
\end{pmatrix}
\begin{pmatrix}
1 & 0 & 0 \\
0 & 1 & 0 \\
3 & 4 & 8 \\
\end{pmatrix}
\begin{pmatrix}
1 & 0 & 0 \\
0 & 1 & 0 \\
4 & 4 & 8 \\
\end{pmatrix}
\begin{pmatrix}
1 & 0 & 0 \\
0 & 2 & 0 \\
0 & 0 & 4 \\
\end{pmatrix}
\begin{pmatrix}
1 & 0 & 0 \\
0 & 2 & 0 \\
0 & 2 & 4 \\
\end{pmatrix}
\begin{pmatrix}
1 & 0 & 0 \\
0 & 2 & 0 \\
1 & 0 & 4 \\
\end{pmatrix}
\begin{pmatrix}
1 & 0 & 0 \\
0 & 2 & 0 \\
1 & 2 & 4 \\
\end{pmatrix}
$ 
\\
$
\begin{pmatrix}
1 & 0 & 0 \\
0 & 2 & 0 \\
2 & 0 & 4 \\
\end{pmatrix}
\begin{pmatrix}
1 & 0 & 0 \\
0 & 2 & 0 \\
2 & 2 & 4 \\
\end{pmatrix}
\begin{pmatrix}
1 & 0 & 0 \\
1 & 2 & 0 \\
0 & 0 & 4 \\
\end{pmatrix}
\begin{pmatrix}
1 & 0 & 0 \\
1 & 2 & 0 \\
1 & 0 & 4 \\
\end{pmatrix}
\begin{pmatrix}
1 & 0 & 0 \\
1 & 2 & 0 \\
2 & 0 & 4 \\
\end{pmatrix}
\begin{pmatrix}
2 & 0 & 0 \\
0 & 2 & 0 \\
0 & 0 & 2 \\
\end{pmatrix}
$ 
\end{tabular} \\

\end{tabular}
\end{ruledtabular}
\end{table*}

\section{Oxygen-deficient perovskite structures on convex hull}

\begin{figure*}[tbp]
\includegraphics[clip,width=\linewidth]{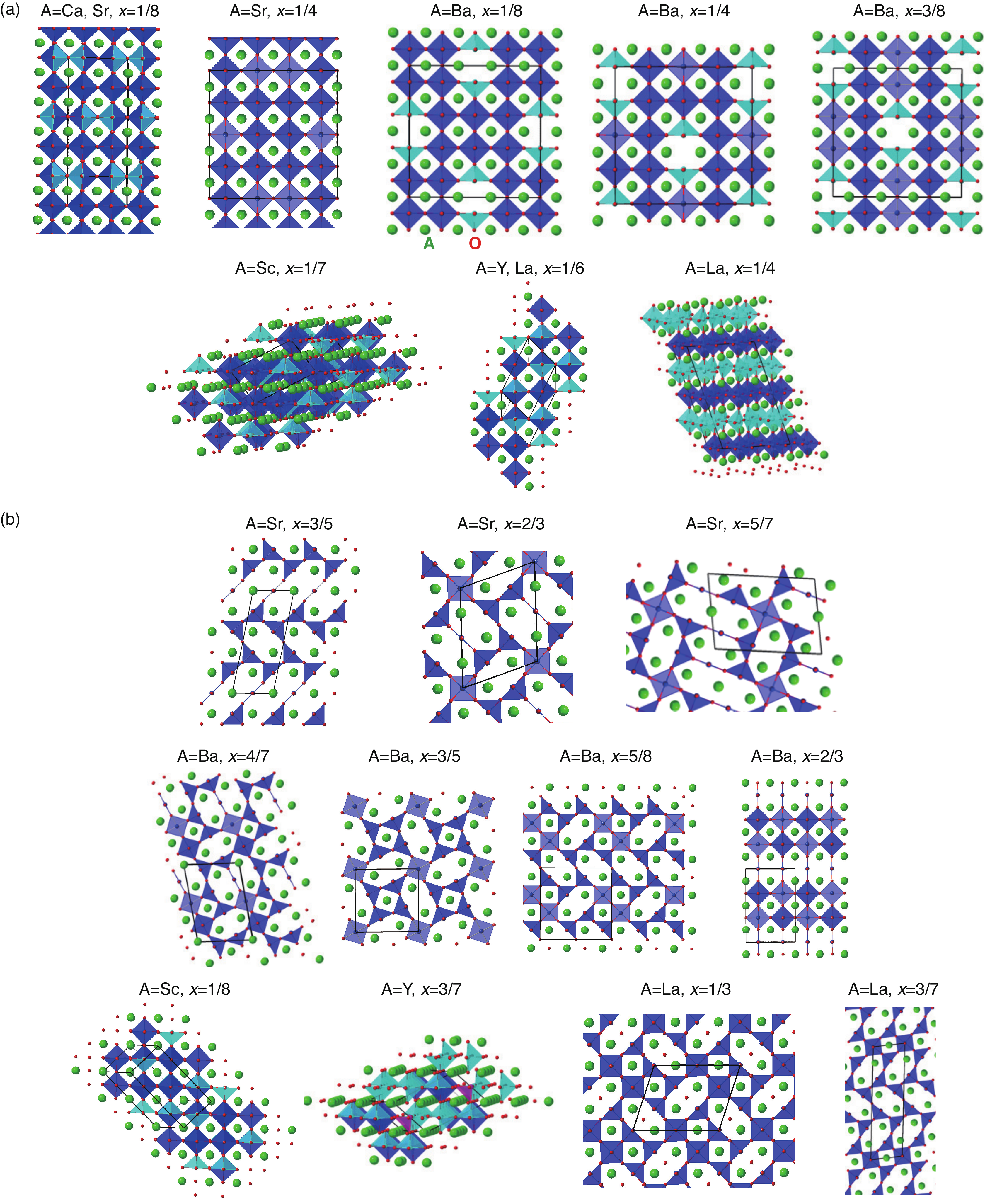}
\caption{
Crystal structures of the oxygen-deficient perovskites (a) on the convex hull and (b) almost on the convex hull at 10 GPa except those shown in Figs. \ref{perovskite:energy-8} and \ref{perovskite:energy-10GPa}.
The dark blue, sky blue, light blue and pink polyhedra show six-fold, five-fold, planer four-fold, tetrahedral four-fold Cu-O polyhedra.
}

\label{perovskite:structure-convex1}
\end{figure*}

Figure \ref{perovskite:structure-convex1} shows the crystal structures of oxygen-deficient perovskites on or almost on the convex hull at 10 GPa.

\bibliography{perovskite}

\end{document}